\documentclass[preprint,12pt]{elsarticle}

\usepackage{amssymb}
\usepackage{amsmath}

\usepackage{lineno}
\usepackage{xcolor}

\journal{Physics Letters A}

\makeatletter
\def\ps@pprintTitle{%
 \let\@oddhead\@empty
 \let\@evenhead\@empty
 \def\@oddfoot{}%
 \let\@evenfoot\@oddfoot}
\makeatother

\begin{document}

\begin{frontmatter}



\title{Compressibility of random walker trajectories on growing networks}


\author{Robert J.\@ H.\@ Ross}
\ead{robert\_ross@hms.harvard.edu}
\author{Charlotte Strandkvist}
\ead{charlotte\_strandkvist@hms.harvard.edu}
\author{Walter Fontana}
\ead{walter\_fontana@hms.harvard.edu}

\address{Department of Systems Biology\\
 Harvard Medical School\\
 200 Longwood Avenue, Boston MA 02115}

\begin{abstract}
We find that the simple coupling of network growth to the position of a random walker on the network generates a traveling wave in the probability distribution of nodes visited by the walker. We argue that the entropy of this probability distribution is bounded as the network size tends to infinity. This means that the growth of a space coupled to a random walker situated in it constrains its dynamics to a set of typical random walker trajectories, and walker trajectories inside the growing space are compressible. 
\end{abstract}


\end{frontmatter}


\section{Introduction}

Numerous physical, biological, cognitive, and social processes are situated in growing (domain-specific) spaces. Developmental processes during embryogenesis and communication networks are obvious examples. In many instances, the growth of the space can profoundly affect the dynamics of the processes it hosts, as has been shown both experimentally and theoretically in the case of cell migration and proliferation \cite{Mort2016,Ross2016b, Ross2017b}. Conversely, processes can shape global characteristics of the space in which they are embedded by determining where growth occurs.  For instance, the ongoing development of the internet is determined by its usage, social networks evolve on the basis of interactions between their constituent individuals, and plasticity in the developing brain is shaped by firing patterns among communicating neurons \citep{Newman2010book,Bear}. 

The question we pursue here is whether the coupling of a process situated inside a growing space, and the indefinite growth of that space, can cause the possible process trajectories to condense into a typical set in the sense of Shannon's source coding theorem, the germinal result describing the limits of data compression \citep{Shannon,Cover2012,Mackay2003}. Such compressibility means that the entropy rate of the random walker trajectory on the network is not divergent. The entropy we consider here is the usual Shannon entropy, $H(X) = -\sum^{n}_{i=1}p_{i}\log p_{i}$, where $X$ is a discrete random variable taking values in $\Omega=\{1,2,\ldots,n\}$ with corresponding probabilities $p_{i\in\Omega}$. That the growth of a state space, typically associated with increasing the uncertainty associated with a process, can also serve to constrain the uncertainty associated with that process, is naturally captured by the source coding theorem and its applications to data compression.

To address this question we use a simple model of a random walker whose position on a network constitutes the attachment point for a new node in a growth event. We show that this coupling suffices for the set of walker trajectories to become compressible, which stands in contrast to three other scenarios in which network growth is decoupled from the walker's position.
Our model is related to an approach by Saram\"aki and others \cite{Saramaki2004, Evans2005, Bloem2018} used in the context of growing scale-free networks, but is a much simplified version. We refer to our model as `walker-induced network growth' or WING \cite{Ross2018b,Ross2018d}.

\section{Results}

In WING, a random walker situated on a network steps from node to node following the edges of the network. At each growth event, whose timing is independent of the walker's motion, one new node is connected with a single edge to the location of the walker. This means that just after addition to the network, new nodes are of degree $k = 1$. We treat the model as a continuous-time Markov chain \cite{Gillespie_orig} in which the times of occurrence of a movement event and of a growth event are exponentially distributed with rates $r_W$ and $r_N$, respectively. There is no limit to the size of the network. In \cite{Ross2018b} we treated the case of multiple self-excluding walkers, but in the present case we will consider only a single walker.

The number of nodes in the network at time $t$ is denoted with $V(t)\in\mathbb{N}$, and the number of edges with $E(t)\in\mathbb{N}$. $V(t)=N(t)+N_0$, where $N(t)$ is the number of growth events that have occurred up to time $t$, and $N_0$ the number of nodes in the seed network.  Throughout this work we typically use a seed network of $N_{0} = 5$, with each initial node connected to all the other initial nodes\footnote{In graph theory notation our seed network is the complete graph, $K_{5}$.}, although this choice does not matter for the results we present. Each node $i$ is uniquely labelled at its creation by the count of growth events that have occurred up to and including its creation. Thus, $i \in \{ 1, 2, ..., V\}$ with the last node labelled $V$. In the case of an initial seed network of five nodes, these initial nodes are simply labelled 1 to 5 in no particular order. The next node added to the network would be labelled 6, and so on. The degree of a node $i$ is denoted by $k_{i}$. 

We compare the behavior of WING with three distinct network growth mechanisms in which the walker plays no role in determining the location at which the network grows. Rather, the random walker only serves as a local observer of the network. These network growth algorithms, while not the main focus of our study, will serve as useful pedagogical tools for contrast with the WING model. In `uniformly random' growth (UR), a new node of degree $k$, whose degree is chosen uniformly at random from the set $\{1,2,\ldots,V\}$, forms one edge to each of $k$ nodes randomly picked without replacement from within the current network. In `fully connected' growth (FC), a new node connects to all nodes in the current network, which therefore remains fully connected. In the Barab\'asi and Albert growth model (BA), at each growth event $n$, a new node is connected to a single node $i$ in the network with a probability $p_i(n)$ proportional to its degree $k_i$, $p_i(n)=k_i/[2E(n)]$ with $n$ indexing the growth event. This is also known as preferential attachment. The probability $p_i(n)$ is the same as the equilibrium probability ($r_{W}\to\infty$) of finding an unbiased random walker at a node of degree $k_i$ in a \emph{non-growing} network. The exact degree distribution generated by the BA model was derived in \citep{Krapivsky2000, Dorogovtsev2000} as $p_{BA}(k) = 4/[k(k+1)(k+2)]$. 

Depending on the values of $r_{W}$ and $r_{N}$, WING generates different network structures \cite{Saramaki2004,Evans2005, Ross2018b,Ross2018d,Cannings2013}. For instance, if $r_{W} = 0$, $r_{N} > 0$, the random walker will not move from its initial position, yielding a network with a `star' structure.  As the ratio $r_{W}/r_{N} \rightarrow \infty$ in the limit, the probability of finding the walker at node $i$ will become the equilibrium distribution for a non-growing network, thus yielding a network with the BA degree distribution $p_{BA}(k)$. For $0 < r_{W} < \infty$, and fixed $r_{N}$, networks have degree distributions in between these extremes. Importantly, the degree distributions generated by WING with a single walker become rapidly stationary for any fixed values of $r_{W}$ and $r_{N}$ \citep{Ross2018b}. In UR, FC, and BA, the network structure does, by definition, not depend on the motility of the walker.

\begin{figure}
\centering
\includegraphics{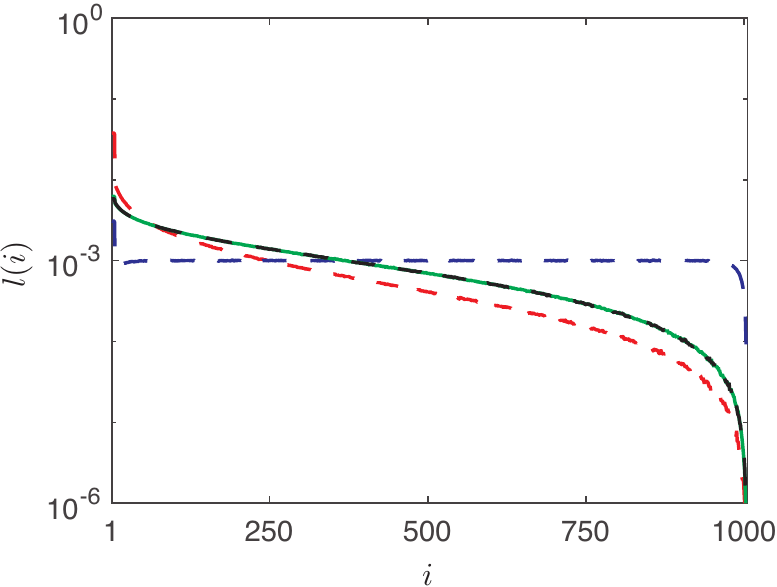}
\caption{The frequency of node labels in a trajectory, calculated using (\ref{eq:visit}), depends on the network growth mechanism:  WING (blue dashed), BA (red dashed), UR (green dashed), and FC (black dashed). FC and UR network growth appear to generate the same distribution. $r_{W} = 1$, $r_{N}=1$, $N_{0} = 5$, $N = 1000$, $V = 1005$, averaged over $R=100,000$ replicates for each growth mechanism.}
\label{fig:labelvisits}
\end{figure}

To obtain a sense for the differences in behavior of a random walker on networks grown with different network growth algorithms, we collect a sample of $R$ trajectories of the random walker, where $R$ is the total number of simulation replicates, for different network growth algorithms. Each trajectory $\tau_{r}$ is a string of node labels $\tau_{r,n}$, where $r$ indicates the replicate number and $n$ indexes the growth event in that replicate. 
For example, given the $r$th trajectory is: $$\tau_{r}=(2,3,6,\fbox{3},4,\ldots),$$ $ \tau_{r,4}$ (indicated by the box), refers to the fourth growth event, which occurred when the walker was situated on the node with label 3, $\tau_{r,4}=3$. This means the walker was situated at the third oldest node in the network when the fourth growth event occurred, in simulation replicate $r$. Similarly, $\tau_{r,5}=4$ means the walker was situated at the fourth oldest node in the network when the fifth growth event occurred, while $\tau_{r,1}=2$ means the walker was situated at the second oldest node in the network when the first growth event occurred, and so on. We provide a figure in the Supplementary Material to further demonstrate how the trajectory of the random walker is recorded \citep{SM}. The probability of the walker being at node labelled $i$ when the $n$th growth event happens, and the probability of the walker being at node $i$ for any growth event are therefore respectively:
\begin{align}
    l(i,n) &= \dfrac{1}{R}\sum_{r=1}^R\delta(i-\tau_{r,n}), \label{eq:track}\\
    l(i) &= \dfrac{1}{N}\sum_{n=1}^N l(i,n), \label{eq:visit}
\end{align}
where $\delta(x)=1$ if $x=0$ and $\delta(x)=0$ otherwise.

\begin{figure}
\centering
\includegraphics[scale=0.5]{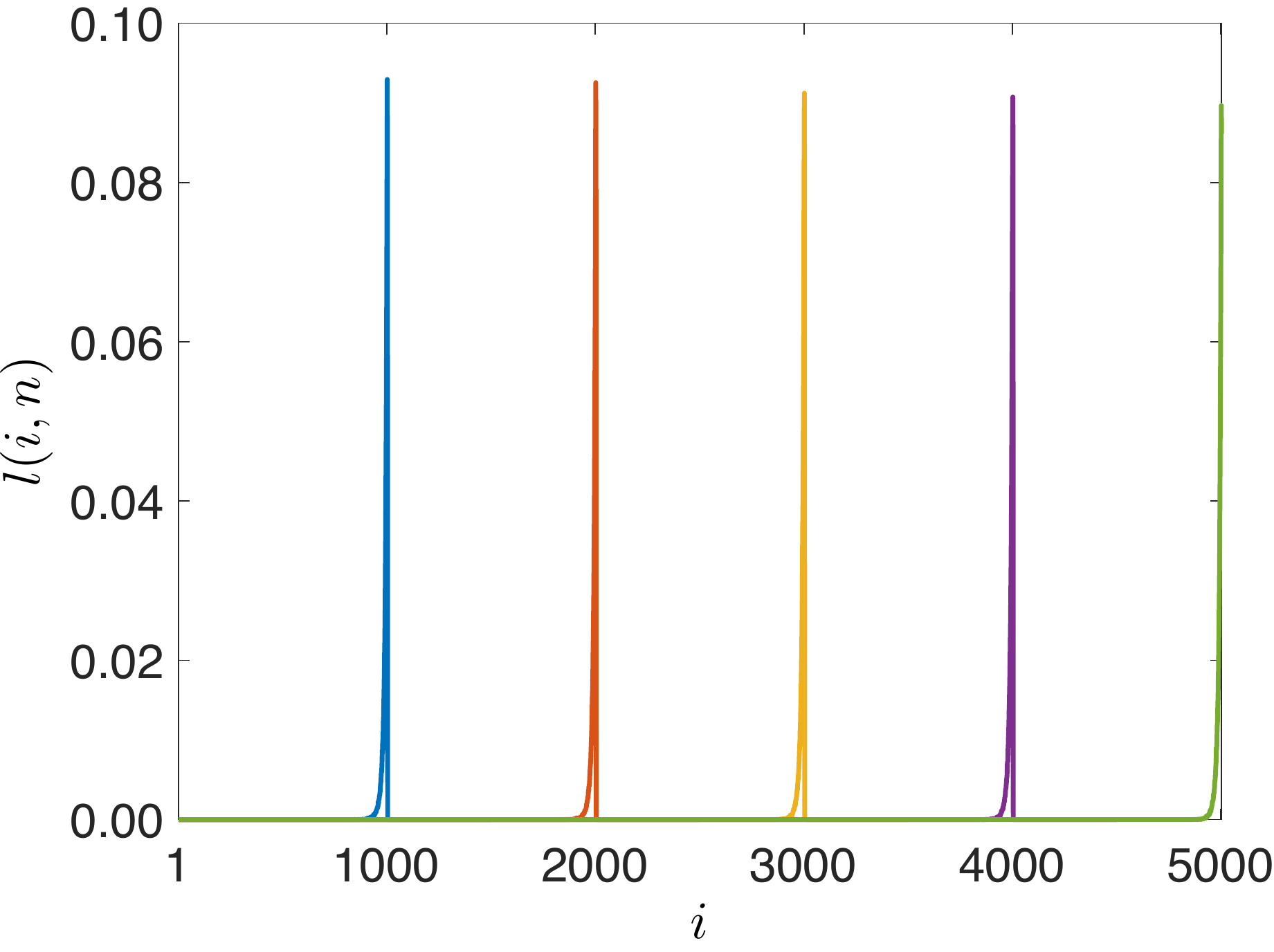}
\caption{The graph shows the distribution $l(i,n)$ described by (\ref{eq:track}) recorded at different growth events or, equivalently, network sizes $V$.  This figure demonstrates that the position of the random walker on a network grown with WING can best be thought of as a traveling wave. The abscissa is the node label and color indicates the growth event, $n$, at which $l(i,n)$ was recorded within the same simulation. $n=1,000-N_{0}$ (blue), $n=2,000-N_{0}$ (red), $n=3,000-N_{0}$ (orange), $n=4,000-N_{0}$ (purple), and $n=5,000-N_{0}$ (green). $r_{W} =1$, $r_{N} = 1$, $N_0=5$.}
\label{fig:wave}
\end{figure}

We begin by comparing the numerically obtained label distribution $l(i)$, as described by (\ref{eq:visit}), for UR, FC, BA and WING. Intuitively, we might expect that the random walker was more likely to be at `older' during growth events than `younger' nodes, simply because many of the younger nodes would not have been part of the network yet. However, Fig.\@ \ref{fig:labelvisits} indicates that for WING, the frequency of any node label appearing in a trajectory is essentially uniform, in marked contrast to the distributions observed for UR, FC, and BA. In UR, FC and BA, the age of a node influences its probability of appearing in a random in the expected way; older nodes appear with higher frequency. The reason for the difference in behavior between UR, FC and BA, and WING is that WING establishes a spatial correlation between the random walker and younger nodes, favoring their inclusion in the random walker trajectory.

Fig.\@ \ref{fig:wave}, shows the distribution (\ref{eq:track}) for WING at progressive times network sizes, providing some clues. At any given network size $V$, the mass of the distribution gravitates around the younger nodes. This tendency of the random walker to visit younger nodes in the network as the network grows, causes a traveling wave in $l(i,n)$.

\begin{figure}
\centering
\includegraphics{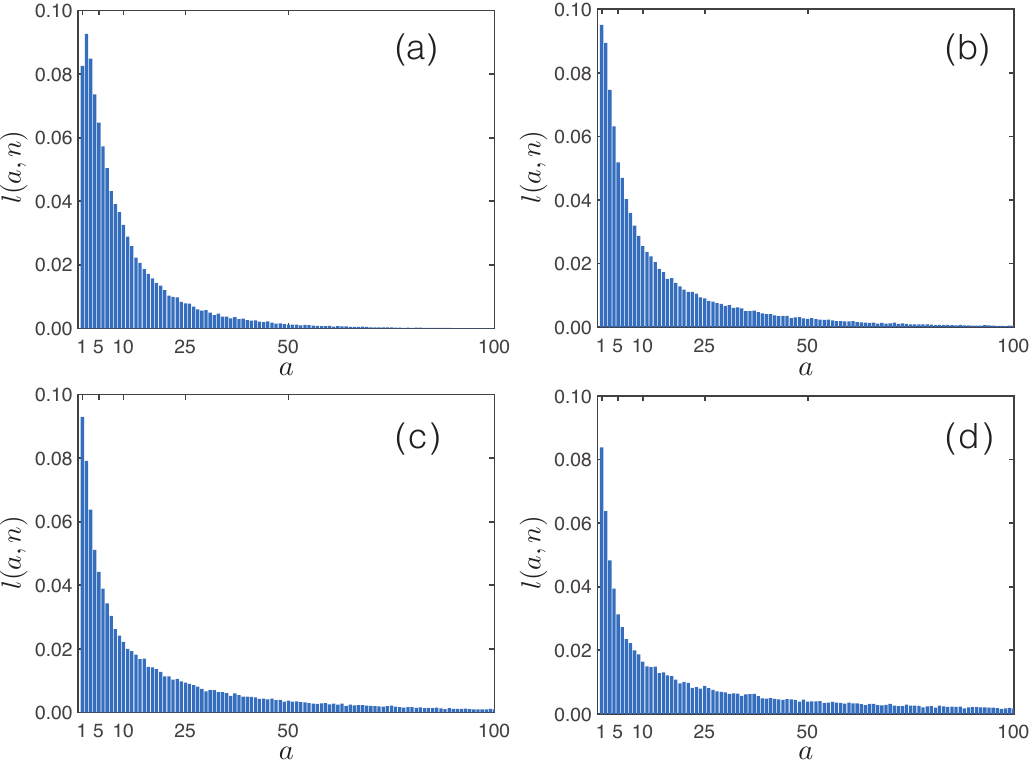}
\caption{The distribution $l(a,n)$, where $a=V-i+1$ is an age label, for different values of $r_{W}$ plotted against node age. (a): $r_{W} =1$; (b): $r_{W} =2$; (c): $r_{W} =3$; (d): $r_{W} =5$. In all panels, $n = 10,000$.}
\label{fig:trackrw}
\end{figure}

To examine $l(i,n)$ at different values of the motility $r_{W}$, it is more intuitive to relabel nodes with respect to age; thus, a node with growth event label $i$ now becomes a node with age label $a=V-i+1$ (age labels of all nodes change with each growth event). This flips each wave front in Fig.\@ \ref{fig:wave} from left to right. Fig.\@ \ref{fig:trackrw} shows that the tail of $l(a,n)$ becomes heavier with increasing $r_W$. This is to be expected, as increasing $r_{W}$ relative to $r_{N}$ counteracts the effect of the spatial correlation between the new node and the random walker. Figures \ref{fig:wave} and \ref{fig:trackrw} suggest that the walker's probability of being at a particular node when the next growth event occurs depends on the age of that node, and is constant regardless of network size. Fig.\@ \ref{fig:trackrw} further suggests that this property holds for any fixed $r_{W}$ and $r_{N}$, although the exact values of $l(a,n)$ depend on the ratio of $r_{W}/r_{N}$.

We next study the entropy rates associated with the random walker trajectories. The entropy rate of a stochastic process is defined as
\begin{align}
H(\mathcal{X}) = \lim_{n \rightarrow \infty} \dfrac{1}{n}H(X_{1},X_{2},X_{3},...,X_{n}),
\label{eq:entropy_rate}
\end{align}
when the limit exists \citep{Cover2012}.  Given the manner in which we generate $l(i,n)$, we should treat each possible random walker trajectory, $(X_{1},X_{2},X_{3},\ldots,X_{n})$, as a series of dependent random variables being sampled from a growing probability distribution. Because it is not computationally feasible to obtain this distribution, we treat each $X_{i}$ as if it were an independent random variable sampled from a growing probability distribution. Assuming independence maximizes the entropy rate associated with repeated sampling from a random variable \citep{Cover2012}, yielding an upper bound for the entropy rate of our trajectories as
\begin{align}
H(\mathcal{X}) \leq \lim_{N \rightarrow \infty} \dfrac{1}{N}\sum_{n=1}^{N}H(X_{n}).
\label{eq:entropy_rate_ind}
\end{align}
In (\ref{eq:entropy_rate_ind}), each $X_{n}$ is a discrete random variable with outcomes in $\{1,2,\dots,V\}$. The total number of vertices in the network includes the size, $N_0$, of the seed network. Thus, $X_{1}$ is the first growth event and so has $N_{0}$ outcomes, each associated with a probability. Similarly, $X_{5}$ is the fifth growth event and so has $N_{0}+4$ outcomes, each associated with a probability.

\begin{figure}
\centering
\includegraphics{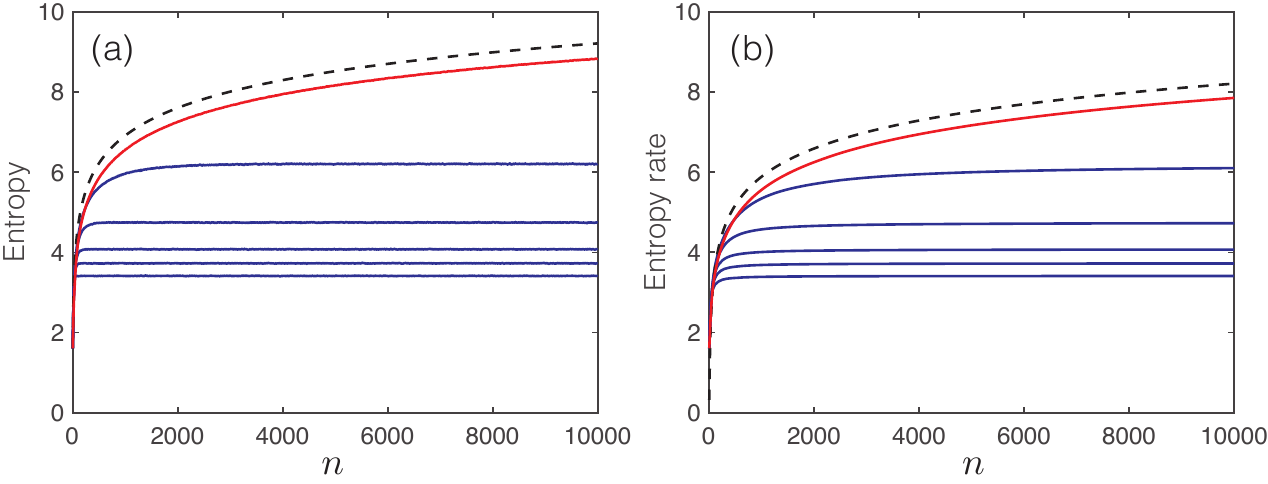}
\caption{(a): The Shannon entropy $-\sum^{V}_{i=1}l(i,n)\log l(i,n)$ of $l(i,n)$ as defined in (\ref{eq:track}) is graphed against growth $n$. (b): The evolution of the bound to the entropy rate, as defined by (\ref{eq:entropy_rate_ind}), of $l(i,n)$. Growth mechanisms: FC (black dashed), BA (solid red), and WING (solid blue) for different values of $r_{W}$.  In both panels, the greater $r_{W}$, the greater the entropy. Bottom to top: $r_{W} = 1,\,2,\,3,\,5,\,10$.}
\label{fig:entropy}
\end{figure}

For FC it is possible to calculate the entropy rate directly with certain assumptions. If we treat the position of the random walker as effectively in equilibrium over the network, meaning $r_{W} \gg r_{N}$, we have that $p_{i} \approx 1/V(t)$. The trajectory of a random walker on a network that grows under FC can be thought of as sampling from a fair $(N(t)+N_0)$-sided die. The entropy rate is therefore
\begin{align}
\lim_{N \to \infty}\dfrac{1}{N}\log{\dfrac{N!}{(N_{0}-1)!}},
\end{align}
which diverges as $N \rightarrow \infty$. A growing fair die can be seen as the `most' divergent entropy rate for the network growth mechanisms we discuss here. Fig.\@ \ref{fig:entropy} shows that, for WING, finite entropy rates appear to exist for all values of $r_{W}$ and are an increasing function of $r_{W}$. In contrast, the entropy rates for BA and FC (and UR) appear divergent.

We next aim to determine whether the distribution described by (\ref{eq:track}) for WING converges on a distribution with finite entropy in the limit $V\to\infty$. If so, the entropy rate associated with the trajectories of the random walker will be well-defined and the trajectories compressible. To demonstrate that $l(i,n)$ has finite entropy as $V \to \infty$ in the limit, we proceed in two parts. First, we demonstrate that the probability of being at a node of a given age when the next growth event occurs is bounded below as $V \to \infty$.  If this were not the case, the entropy of the distribution would not be bounded as with FC. Second, we argue that the tail of $l(i,n)$ decays appropriately so as to admit a finite entropy asymptotically.  For instance, if the tail decayed with $O(a^{-1})$ it would not admit a finite entropy.

To approach the first condition, we again recast $l(i,n)$ in terms of node age $a=V-i+1$, as in Fig.\@ \ref{fig:trackrw}, and write $\rho(a)$ for $l(a,n)$ as $V \to \infty$. The claim is that
\begin{align}
\rho(a) > \alpha_{a}, \ \forall \ a \ \text{as} \ V \to \infty, 
\label{eq:alpha}
\end{align}
where $\alpha_{a}$ is a constant dependent on the age of the node and the parameters of the model. In this notation, $a=1$ is the last node that was added to the network, $a=2$ is the second to last node added to the network, and so on. Expression (\ref{eq:alpha}) states that the probability of being at a node of age $a$ when a growth event occurs is bounded below in the limit of large network size. 

We show first that this is true, in particular, for the probability $\rho(1)$ of being at the newest node in the network when the next growth event occurs. Let $p_W(k)$ denote the probability that the walker is situated at a node of degree $k$ when a growth event occurs. We define
\begin{align}
\kappa = \sum_{k=1}^{\infty}\frac{p_{W}(k)}{k+1},
\label{eq:kappa}
\end{align}
where $\kappa$ is the probability that if the walker were to move following a growth event it would select the new node.
The next event in the system is either a step by the walker or another growth event. The former occurs with probability $m=r_W/(r_W+r_N)$; the latter with probability $g=r_N/(r_W+r_N)$. Thus, the probability that the walker is at the node of age $1$ when the next growth event occurs is at least the probability of the event sequence `one step in the right direction, followed by a growth event':
\begin{align}
\rho(1) > m\kappa g.
\label{eq:boundfor1}
\end{align}
It is at least this probability, because there are many, more circuitous, paths to reach the new node before the next growth event. If $\kappa$ is stationary in the limit, the bound (\ref{eq:boundfor1}) is independent of network size and thus of time. Analogous reasoning leads to
\begin{align}
\rho(a) > m\kappa g^{a}.
\label{eq:boundforall}
\end{align}
The claim (\ref{eq:alpha}) is shown if we can ascertain that $\kappa$ has properties that make it a meaningful proxy of local network structure. For example, $p_{W}(1)$ has a lower bound greater than zero. In other work we showed that this property does indeed hold \citep{Ross2018b}.

Implicit in the above is that all higher order probabilities can be bounded below in a similar manner. For instance, if we define $\rho(a,b)$ as the joint probability that a growth event occurs at a node of age $b$ followed by a growth event at a node of age $a$, then for $a \leq b$
\begin{align}
\rho(a,b) > \rho(b-a+1)g^{a-1}\,\dfrac{m}{b+1}g.
\label{eq:higher}
\end{align}
The right-hand-side of (\ref{eq:higher}) describes a growth event at a node of age $b-a+1$, followed by a further $a-1$ growth events at this node. The node the walker is situated on has now age $b+1$ and has at most $b+1$ neighbors, one of which has age $a$, and so a movement event to the node of age $a$, followed by a growth event, completes (\ref{eq:higher}). Similar bounds exist for joint probabilities in which $a>b$.
It follows from these bounds that conditional probabilities, such as $\rho(a|b)$, defined as a growth event at the node of age $a$ conditioned on the preceding growth event having occurred at the node of age $b$, are also bounded below, since $\rho(a|b)=\rho(a,b)/\rho(b)$ and the ratio is bounded.

FC is an example of a growth mechanism for which $\rho(a)$ is not bounded below in the limit. Although the new node is always adjacent to the node at which the walker is situated, the probability of moving to this new node (or any other neighboring node), given a movement event, is $1/(V-1)$, which is not bounded below in the limit $V \rightarrow \infty$. Similarly, for BA network growth, the new node is increasingly less likely to be adjacent to the walker as $V \rightarrow \infty$, and so the distance from the walker to the new node is a function of network size.

Continuing with this reasoning, we can calculate the expected degree $\langle k_{i} \rangle$ of a node given its age in the limit $V\to \infty$.  Each node introduced to the network is of degree $k =1$, and has a probability $\rho(1)$ of the walker being situated on it at the next growth event and so having its degree increased by one; or a probability $\rho(2)$ of the walker being situated on it at the following growth event and so having its degree increased by one; and so on. Therefore,
\begin{align}
\langle k_{i} \rangle = 1 + \sum_{j=1}^{\infty}1 \cdot \rho(j) = 2,
\label{eq:av_degree}
\end{align}
asymptotically. The independence of $\langle k_{i} \rangle$ from age was suggested already by Fig.\@ \ref{fig:labelvisits}.

Finally, we argue that the tail of the probability distribution $\rho(a)$ decays in such a way that the entropy is finite as $V \rightarrow \infty$. After a growth event has occurred at a node of unknown age $\hat{a}$ (and no event has occurred since), the random walker must be situated on a node that shares an edge with the newest node in the network. Hence, the expected distance between the node $\hat{a}$ at which the walker is located and a node of age $a$ is $d(\hat{a},a;V) = d(1,a;V) - 1 \approx d(1,a;V)$.  
In the Supplemental Material \cite{SM} we numerically demonstrate that the expected distance between two nodes appears to be a linear function of the difference in their age (independent of $V$). If we assume that the expected distance between node 1 and node $a$ grows linearly with $a$, we can think of the problem as diffusion of a random walker in one dimension. Therefore, the probability the next growth event occurs at a node of age $a$ is bounded above by $O(m^{a})$ as $V \rightarrow \infty$, and so $\rho(a)$ would admit a finite entropy. Alternatively, if the expected distance between nodes in the network does not increase as a linear function of age difference (but logarithmically, for instance), it is possible that despite lower bounds, $\rho(a)$ would not admit a finite entropy as $V \to \infty$.

It is well understood that compressibility (in practice) takes advantage of non-uniformities in the probability distribution being sampled from \cite{Shannon, Cover2012, Mackay2003}. We have argued that WING generates a nonuniform distribution of the walker's position that has a finite entropy as $V \rightarrow \infty$. Therefore, WING asymptopically generates a typical set of random walker trajectories with the following property
\begin{align}
e^{-N(H(\mathcal{X})+\epsilon)} \leq p(x_{1},x_{2},\ldots,x_{N}) \leq e^{-N(H(\mathcal{X})-\epsilon)},
\end{align} 
where $H(\mathcal{X})$ is the entropy rate associated with the walker's position for fixed values of $0 < r_{W} < \infty$ and $0 < r_{N} < \infty$, and so means the random walker trajectories are compressible.

\section{Conclusion}

These results show that growth, typically associated with increasing the number of states accessible to a process, can nonetheless also function to constrain its likely outcomes.  In WING, the coupling between growth and random walker generates a traveling wave that biases the walker towards occupying the youngest nodes in the network.  We have argued that an asymptotic property of this traveling wave---finite entropy---delivers compressibility. The cost for the compressibility of WING trajectories is that the growth mechanism must know the location of the walker. This is not unlike using informed manipulation to control uncertainty in variations of Maxwell's demon \cite{Szilard1929, Landauer1961}.

It is an open question whether the compressibility of random walker trajectories is a common property of network growth mechanisms that couple network growth and the position of a random walker (or another process situated on the network). However, the simple setup of our model suggests that the phenomenon might be general.  It is worth highlighting the effect different types of growth may have on compressibility. In this work we have employed linear network growth, but had we employed exponential growth we would not have observed compressibility of WING trajectories, because the effective network growth rate is $r_N V$, and the ratio $r_W/(r_N V)$ tends to zero as $V \to \infty$ \citep{Ross2018b}. As a result, the local environment of the walker becomes increasingly `star-like', and so differs from the situation in which $r_{W} = 0$, when random walker trajectories are trivially compressible.

Finally, the Shannon entropy is the value of the R\'enyi entropy $H_{\alpha}(X) = 1/(1-\alpha)\log\sum^{n}_{i=1} p^{\alpha}_{i}$ as $\alpha \to 1$. Other limiting values of the R\'enyi entropy have attracted interest. Specifically, $H_{1/2}(X) = 2\log\sum^{n}_{i=1} p^{1/2}_{i}$ has been shown to bear relation, in the limit $n \to \infty$, to the average guesswork needed for identifying the value of a random variable \cite{Massey1994, Arikan1996}. Following our treatment of entropy rates in WING, $H_{1/2}(X)$ would converge. This means that the expected number of guesses to find the walker (using an optimal guessing strategy) would remain finite (and bounded) as $V \to \infty$. Being able to efficiently find the walker in the network is important for the practical feasibility of WING and, more generally, the feasibility of any network growth mechanism whereby a process on the network must be analyzed in some manner before the growth event can occur.  

\section{Acknowledgements}

RJHR would like to acknowledge Ioana Cristescu, Muriel M\'edard and Ken Duffy for helpful discussions related to the work presented here.






\end{document}


\begin{center}
{\large Compressibility of random walker trajectories on
growing networks}\\[0.2cm]
{\large Supplemental Material}

\vspace*{1cm}
Robert J.\@ H.\@ Ross, Charlotte Strandkvist, Walter Fontana
\end{center}


\vspace*{1cm}

\setcounter{section}{0}
\setcounter{equation}{0}
\setcounter{figure}{0}
\setcounter{table}{0}
\setcounter{page}{1}
\makeatletter
\renewcommand{\theequation}{S\arabic{equation}}
\renewcommand{\thefigure}{S\arabic{figure}}
\renewcommand{\bibnumfmt}[1]{[S#1]}
\renewcommand{\citenumfont}[1]{S#1}
\renewcommand{\thesection}{\Alph{section}}

\renewcommand\thefigure{SF\arabic{figure}}    
\setcounter{figure}{0}    
\setcounter{section}{0}
\renewcommand{\thesection}{\Alph{section}}

\section{How the random walker trajectory, $\tau_{r}$, is recorded}

\begin{figure}[h!]
\centering
\includegraphics[scale=0.4]{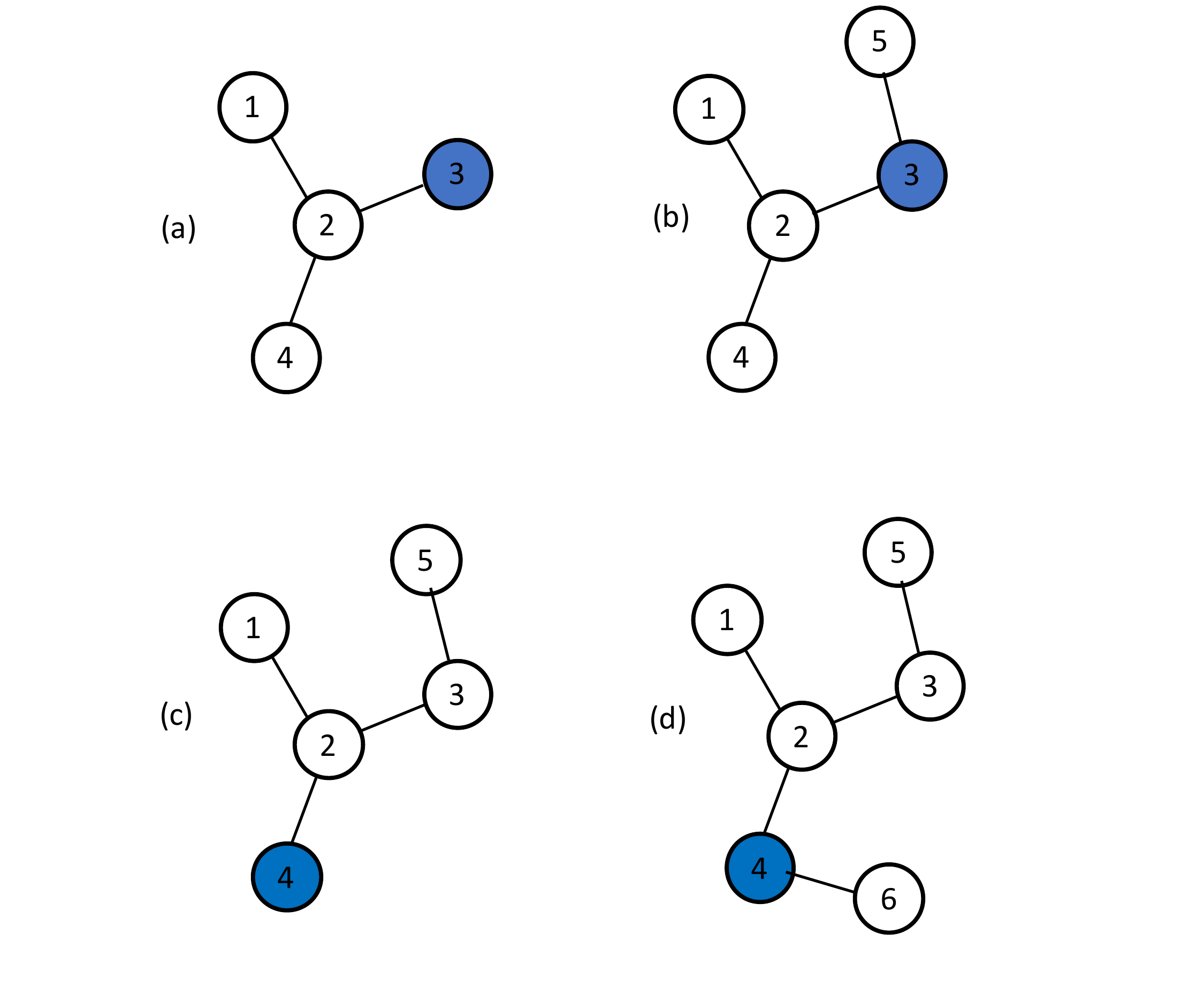}
\caption[How the random walker trajectory, $\tau_{r}$, is recorded]{In this example the network growth algorithm is WING. The position of the random walker is denoted by blue. Our initial seed network is the network denoted by (a), and so no growth events have occurred yet. In (a) the node labels, 1 to 4, have been randomly assigned to the four initial nodes in the network.  In (b), a growth event has occurred while the random walker is situated at the node labelled 3, which means we record $\tau_{r,1} = 3$, because the random walker was at the node labelled 3 when the growth event occurred. In (c), the random walker now occupies the node with label 4.  In (d), a growth event has occurred while the random walker is situated at the node labelled 4, which means we record $\tau_{r,2} = 4$.  This process is continued until we reach the desired $n$ growth events.  This is one random walker trajectory $\tau_{r}$.}
\label{suppfig:tau}
\end{figure}

\newpage

\section{Local and global degree distribution of WING}

Let $p_W(k)$ denote the probability that the walker is situated at a node of degree $k$ when a growth event occurs. To obtain $p_W(k)$ numerically we record the trace $t_r$ of degrees seen by the walker in a simulation replicate $r$, collect the frequency with which the degree is $k$ at growth event $n$ across replicate traces $t_r$ ($r$ indexing the replicate) each comprising $N$ growth events, and average over $n$. Denoting the degree the walker observes at event $n$ of trace $t_r$ by $t_{r,n}$, we have
\begin{align*}
    p_W(k,n) &= \dfrac{1}{R}\sum_{r=1}^R\delta(k-t_{r,n}) \\
    p_W(k) &= \dfrac{1}{N}\sum_{n=1}^V p_W(k,n)
\end{align*}
where $\delta(x)=1$ if $x=0$ and $\delta(x)=0$ otherwise. The global degree distribution $p(k)$ is computed likewise, but instead of observing a single node at growth event $g$, we observe all nodes in the network.

\begin{figure}[h!]
\includegraphics[scale=0.65]{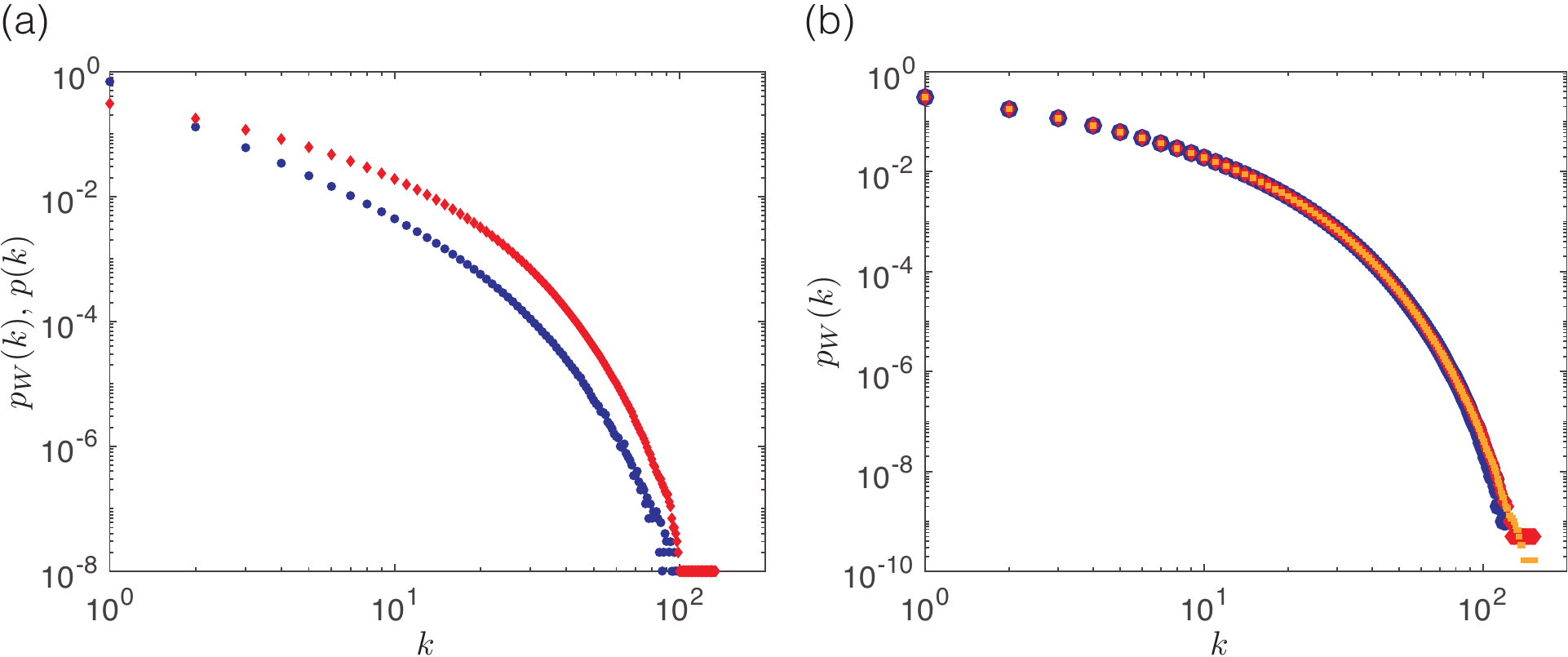}
\caption[Local and global degree distribution]{Local and global degree distribution. Panel (a): The global degree distribution $p(k)$ (blue disks) is compared with the degree distribution seen by the walker $p_W(k)$ (red diamonds) after $N = 10,000$ growth events. $r_W=5$, $r_N=1$. Panel (b): Convergence of $p_W(k)$ to a stationary distribution is rapid. $p_W(k)$ is depicted for different growth extents: $N=10,000$ (blue filled circles), $N=20,000$ (red diamonds), and $N=100,000$ (orange squares).}
\label{suppfig:local}
\end{figure}

\newpage
\section{The average distance between two nodes as a function of age}

\begin{figure}[h!]
\includegraphics[scale=0.65]{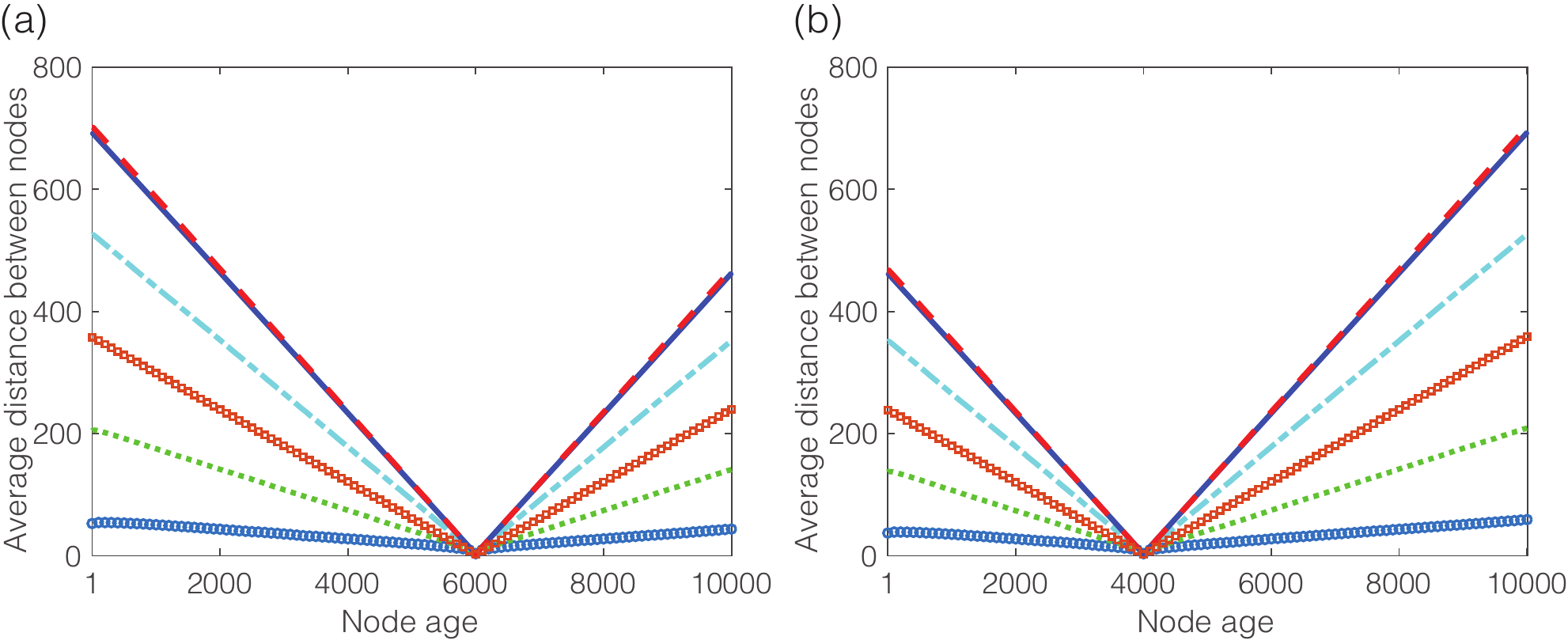}
\caption[Distance between nodes as a function of age]{Distance between nodes as a function of age. Panel (a): The average distance between a node of age $4000$ and all other nodes in a network generated with WING is shown as a function of their age difference for different values of walker motility. $r_W = 10$ (blue circles), $r_W = 5$ (green dashed), $r_W = 2$ (blue dot-dashed), $r_W = 1$ (blue solid), $r_W = 0.5$ (red dashed) and $r_W = 0.1$ (red squares). Panel (b): As in panel (a), but for node $6000$. In both panels $V=10,000$.}
\label{suppfig:distance}
\end{figure}

The average distance between two nodes in networks with WING is a linear function in the difference in their ages and does not depend on the size of the network.  The distance between nodes for networks grown with WING is maximized between $r_{W} = 0.5$ and $r_{W} = 1$. Figure \ref{suppfig:distance} also evidences that the age of a node and its degree do not positively correlate in WING.

